\documentstyle[12pt]{article}
\title{Vortex vs spinning  string:  Iordanskii
force and gravitational Aharonov-Bohm effect.}
\author{G.E. Volovik\\
Low Temperature Laboratory,
Helsinki University of Technology\\
P.O.Box 2200, FIN-02015 HUT, Finland\\
and\\
L.D. Landau Institute for Theoretical Physics,
 Moscow\\
}
\begin{document}
\maketitle
\begin{abstract}
{ We discuss the transverse force acting on the spinning cosmic string, moving
in the background matter. It comes from the gravitational Aharonov-Bohm
effect and corresponds to the Iordanskii force acting on the vortex in
superfluids, when the vortex moves with respect to the
normal component of the liquid.
 }
\end{abstract}

{\it Introduction.}

In superfluids, with their two-fluid hydrodynamics (for superfluid and normal
components of the liquid) there are 3 different topological contributions
to the
force acting on the quantized vortex \cite{3Forces}. The
more familiar Magnus force arises when the vortex moves with respect to the
superfluid vacuum. For the relativistic cosmic string such force is absent
since
the corresponding superfluid density of the quantum physical vacuum is zero.
However the analog of this force appears if the cosmic string  moves in the
uniform background charge density
\cite{DavisShellard,Lee}. The other two forces of topological origin also have
analogs for the cosmic strings: one of them comes from the analog of the axial
anomaly in the core of electroweak string (see  Reviews
\cite{AxialAnomaly}), and
another one -- the Iordanskii force -- is now under active discussion in
condensed matter community \cite{SoninNew,Wexler,Shelankov}.

As distinct from the Magnus force, the Iordanskii force
\cite{Iordanskii,Sonin1}  arises when the vortex moves with respect to the heat
bath represented by the normal component of the liquid, which consists of the
quasiparticle excitations. The latter corresponds to the matter in particle
physics.  The interaction of quasiparticles with the velocity field of the
vortex
resembles the interaction of the matter with the gravitational field
induced by such cosmic string, which has an angular momentum, -- the so-called
spinning cosmic string
\cite{CausalityViolation}. The spinning string induces the peculiar space-time
metric, which leads to the time delay for any particle orbiting around the
string
with the same speed, but in opposite directions \cite{Harari}. This gives rise
to the quantum gravitational Aharonov-Bohm effect
\cite{CausalityViolation,MazurComment1,MazurComment2}.
We discuss here how the same effect leads to the asymmetry in the scattering of
particles on the spinning string and finally to the Iordanskii lifting force
acting on the spinning string.

{\it Vortex vs spinning cosmic string.}

To clarify the analogy between  the Iordanskii force and AB effect,
let us consider the simplest case of phonons propagating in the velocity field
of the quantized vortex in the Bose superfluid $^4$He.  According to the Landau
theory of superfluidity, the energy of quasiparticle moving in the superfluid
velocity  field
${\bf v}_s({\bf r})$ is  Doppler shifted: $E({\bf p})= \epsilon({\bf p})+  {\bf
p}\cdot{\bf v}_s({\bf r})$. In the case of the phonons with the spectrum
$\epsilon({\bf p})=cp$, where $c$ is the sound velocity,  the energy-momentum
relation is thus
\begin{equation}
\left(E- {\bf p}\cdot{\bf v}_s({\bf r})\right)^2=c^2p^2
 ~.
\label{PhononSpectrum}
\end{equation}
 The Eq.(\ref{PhononSpectrum}) can be written
in the general Lorentzian form with  $p_{\mu}=(E,{\bf p})$:
\begin{equation}
g^{\mu\nu}p_\mu p_\nu=0 ~~,~~
g^{00}=1,~~~~g^{0i}=-v_{s}^i,~~
g^{ik}=- c^2 \delta^{ik} +v_s^iv_s^k ~~.
\label{gikVortex}
\end{equation}
 Thus the dynamics of phonons in the presence of
the velocity field is the same as the dynamics of photons in the gravity field
\cite{Unruh1}: both  are described by the lgiht-cone equation $ds=0$,
where the interval $ds$ is given by the inverse metric $g_{\mu\nu}$:
\begin{equation}
 ds^2=g_{\mu\nu}dx^\mu dx^\nu  ~~.
\label{IntervalGeneral}
\end{equation}
Here we are interested in the velocity field circulating around quantized
vortex,
${\bf v}_s=N\kappa\hat{\bf \phi}/2\pi r$, where $\kappa$ is the quantum of
circulation and $N$ is the circulation quantum number. This flow
induces the effective space, where the phonon is
propagating along geodesic curves, with the interval
\begin{equation}
  ds^2=\left(1-{v_s^2\over c^2}  \right)\left(dt +{N\kappa
d\phi\over 2\pi( c^2-v_s^2)} \right)^2
-{dr^2\over c^2}-{dz^2\over c^2}
-{ r^2d\phi^2\over c^2-v_s^2}
 ~.
\label{PhononInterval}
\end{equation}

Far from the vortex, where $v_s^2/c^2$ is small and can be neglected, one has
\begin{equation}
ds^2=\left( dt + {d\phi\over \omega}\right)^2 -{1\over c^2}(dz^2+ dr^2
+r^2d\phi^2)~~,~~\omega={2\pi  c^2 \over  N \kappa}
\label{IntervalVortexAsymp}
\end{equation}
The connection between the time and the azimuthal angle $\phi$  in
the interval suggests that there is a charactiristic angular
velocity $\omega$. For the vortex in superfluid $^4$He, where $\kappa=2\pi
\hbar/m_4$ and $m_4$ is the mass of $^4$He atom, it is
$\omega= m_4 c^2/N\hbar $.  The similar
metric with rotation was obtained for the so-called spinning
cosmic string in $3+1$ space-time, which has  the rotational angular momentum
$J$ concentrated in the string core, and for the spinning particle in the 2+1
gravity
\cite{CausalityViolation,MazurComment2,Staruszkievicz,Deser}:
\begin{equation}
ds^2=\left( dt + {d\phi\over \omega}\right)^2 -{1\over c^2}(dz^2+ dr^2
+r^2d\phi^2)~~,~~\omega={1\over 4JG}
\label{IntervalSpinningStringAsymp}
\end{equation}
where $G$ is the gravitational constant.
This gives the following correspondence between the circulation $N\kappa$
around the vortex and the angular momentum $J$ of the spinning string
\begin{equation}
\kappa N = 8\pi JG~.
\label{JvsN}
\end{equation}
Though we consider the analogy between the spinning string and
vortices in superfluid $^4$He, there is a
general statement that vortices in any
superfluids have the properties of the spinning cosmic strings
\cite{DavisShellard}. In particular, the spinning string generates the density
of the angular momentum in the vacuum outside the string
\cite{JensenKucera}. The density
of the angular momentum in the superfluid vacuum outside the vortex is also
nonzero and equals $\hbar N n$, where $n$ is the density of elementary bosons
in superfluid vacuum:
the density of $^4$He atom in superfluid $^4$He or of Cooper pairs in
superfluid $^3$He.

 {\it Gravitational Aharonov-Bohm effect.}

The   effect peculiar for the spinning string, which can be modelled in
condensed matter, is  the gravitational AB topological effect
\cite{CausalityViolation}. On the classical level the propagation of the
phonons is described by the equation $ds^2=0$. Outside the string the
metric, which enters the interval $ds$, is locally flat. But there is the time
difference for the particles propagating  around the spinning string  in
the opposite directions. For the vortex (at large distances from the core)
this time delay approaches \cite{Harari}
\begin{equation}
2\tau={4\pi \over \omega}~~.
\label{Time Delay}
\end{equation}
This asymmetry between the particles moving on different sides of the vortex is
the origin of the  Iordanskii force acting on the vortex in the presence
of the net momentum  of the quasiparticles. On the quantum level, the
connection between the time variable $t$ and the angle variable $\phi$ in
the metric Eq.(\ref{IntervalSpinningStringAsymp}) implies that the
scattering cross section of phonons (photons) on the vortex should be the
periodic function of the energy with the period equal to   $\hbar \omega$.
The asymmetric part of this cross section gives rise to the  Iordanskii force.

There was an  extreme interpretation of the  gravitational AB
effect put forward by Mazur \cite{CausalityViolation}. He
argued that for the infinitely thin  spinning cosmic string there is a
region where the causality is violated. To avoid causality violation the
string should be  transparent for the excitations and this is possible only if
in the presence of the spinning cosmic string the energy of the elementary
particles is strictly quantized:
$E=n\hbar\omega$. In other words the  gravitational AB effect leads to
the energy quantization in the same manner as the quantization of the
electric charge should take place in the presence of the Dirac magnetic
monopole.  On the other hand there are solutions for cosmic strings,
which do not contain the closed timelike curves \cite{JensenSoleng}. In this
case the severe energy quantization is not necessary, see however
discussion in \cite{MazurComment1,MazurComment2,Glushchenko,Kharkov}. In any
case the periodicity with the period
$\Delta E=\hbar\omega$ is retained and the symmetric part of the scattering
cross section of particle with  energy
$E$ in the background of spinning string with zero mass is
\cite{MazurComment1,MazurComment2}:
\begin{equation}
{d \sigma\over d\theta} = {\hbar c\over 2\pi E \sin^2(\theta/2)}~  \sin^2
{\pi
E\over
\hbar \omega}~.
 \label{DiffCrossSectionString}
\end{equation}

We argue that in addition to this symmetric part there is the topological
asymmetric contribution, which gives rise to transverse cross section
\begin{equation}
\sigma_\perp =\int_0^{2\pi} d\theta ~\sin\theta ~|a(\theta)|^2
\label{sigmaPerpGeneral}
\end{equation}
The asymmetry in the scattering of the
quasiparticles on the velocity field of the vortex has been calculated by Sonin
for phonons and rotons in $^4$He
\cite{SoninNew} and by Cleary \cite{Cleary} for the
Bogoliubov-Nambu quasiparticles in conventional superconductors.
In the case of phonons the propagation is described by the Lorentzian equation
for the scalar field, $g^{\mu\nu}\partial_\mu\partial_\nu ~\Phi=0$.
In the asymptotoc region the quadratic terms ${\bf v}_s^2/c^2$ can be neglected
and this equation can be rewritten as \cite{SoninNew}
\begin{equation}
E^2\Phi -c^2\left(-i\nabla + {E\over c}{\bf v}_s({\bf r})\right)^2\Phi=0 ~~.
\label{ModifiedScalarField}
\end{equation}
This equation maps the problem under discussion to the Aharonov-Bohm (AB)
problem for the magnetic flux tube \cite{AB} with the vector potential ${\bf
A}={\bf v}_s$, where the electric charge $e$ is substituted by the mass $E/c^2$
of the particle \cite{MazurComment1,JensenKucera,Galtsov}.  Because of the
mapping between the electric charge and the mass of the particle, the Lorentz
force, which acts on the flux tube in the presence of electric current, has its
counterpart -- the Iordanskii force, which acts on the vortex in the presence
of the mass current carried by the normal component
\cite{Shelankov}.

If one directly follows the mapping of the phonon scattering  on vortices
described by Eq.(\ref{ModifiedScalarField}) to the AB scattering, one
obtains the  AB result \cite{AB} for the  symmetric part of the differential
cross section, now written in the form of  Eq.(\ref{DiffCrossSectionString}).
There is a not very important difference, which comes from the
definition of the quasiparticle current:  as
distinct from the charged particles in the AB effect, the current in our
case is
not gauge invariant. As a result the scattering of the phonon with
momentum
$p$ and with the energy $E$ by the vortex is somewhat different
\cite{SoninNew}:
\begin{equation}
{d \sigma\over d\theta} = {\hbar c\over 2\pi E }\cot^2{\theta\over 2} ~
\sin^2
{\pi E\over
\hbar \omega}~.
 \label{DiffCrossSectionVortex}
\end{equation}
The difference between Eq.(\ref{DiffCrossSectionVortex}) and
the AB result Eq.(\ref{DiffCrossSectionString}) is
$(c/ 2\pi E )  ~  \sin^2   (\pi E/ \omega)$, which is independent of the
scattering angle $\theta$ and thus is not important for the singularity
at small scattering angles.
For small $E$ the result in Eq.(\ref{DiffCrossSectionVortex}) was obtained by
Fetter \cite{Fetter}. The generalization of the Fetter result for the
quasiparticles with  arbitrary spectrum
$\epsilon({\bf p})$ (rotons in $^4$He and the Bogoliubov-Nambu fermions in
superconductors) was recently suggested in Ref.\cite{Demircan}: In our
notations
it is $(N\kappa^2 p/8\pi v_G^2) \cot^2(\theta/ 2)$, where $v_G=d\epsilon/dp$ is
the group velocity of quasiparticle.

{\it Asymmetric cross section.}

The Lorentz-type Iordanskii force comes from  the  asymmetric singularity
in the cross section \cite{Shelankov}. This additional
topological term is  determined by the same asymptote of the flow velocity,
which
causes singularity at the small angles in the symmetric cross section. The
asymmetric part of  the differential cross section gives the following
transverse
cross section
\cite{SoninNew}
\begin{equation}
\sigma_\perp ={\hbar \over p} ~ \sin  {2\pi
 E\over \hbar \omega}
\label{sigmaPerpVortex}
\end{equation}
At low $E$ this result was
generalized for arbitrary excitations with the spectrum $E({\bf
p})=\epsilon({\bf
p})+  {\bf p}\cdot{\bf v}_s$ moving in the background of the
velocity field ${\bf v}_s$ around the vortex, using a simple classical
theory of scattering
\cite{SoninNew}. Far from the vortex, where the circulating velocity is
small, the trajectory of the quasiparticle is almost the straight line
parallel, say, to the axis
$y$, with the distance from the vortex line being the impact
parameter
$x$. It moves along this line with the almost constant momentum
$p_y\approx p$ and  almost constant group velocity
$v_G=d\epsilon/dp$. The change in the transverse momentum during
this motion is determined by the Hamiltonian equation $dp_x/dt=-\partial
E/\partial x=-p_y \partial
v_{sy}/\partial x$, or $dp_x/dy=-(p/v_G)\partial
v_{sy}/\partial x$. The transverse cross section is obtained by
integration of $\Delta p_x/p$ over the impact parameter $x$:
\begin{equation}
\sigma_\perp = \int_{-\infty}^{+\infty}{dx\over
v_G}\int_{-\infty}^{+\infty}dy {\partial v_{sy}\over \partial x}
={
N\kappa\over v_G}~.
\label{sigmaPerpLinear}
\end{equation}
Note that this result is a pure classical: the Planck constant $\hbar$
drops out.

{\it Iordanskii force on spinning string.}

 This
asymmetric part of scattering,   which describes the momentum transfer in
the transverse direction, after integration over the  distribution of
excitations  gives rise to the transverse force acting on the  vortex if
the vortex moves with respect to the normal component. This is  the
Iordanskii force:
\begin{equation}
{\bf f}_{\rm Iordanskii}=\int {d^3p\over (2\pi)^3}\sigma_\perp(p) v_G n({\bf
p})
 {\bf p}\times {\hat{\bf z}} =- N\kappa  {\hat{\bf z}}\times  \int {d^3p\over
(2\pi)^3}  n({\bf p})
 {\bf p}
= N\kappa  {\bf P}_n\times {\hat{\bf z}}
\label{IordanskiiForce}
\end{equation}
It depends only on the density of mass current ${\bf
P}_n$ carried by excitations (matter) and on the circulation
$N\kappa$ around the vortex. This confirms the topological origin of this
force. In the case of the equilibrium distribution of quasiparticles one has
${\bf P}_n=\rho_n{\bf v}_n$, where $\rho_n$ and ${\bf v}_n$ are the density
and velocity of the  normal component of the liquid (to avoid the conventional
Magnus force, we assume that the asymptotic velocity of the superfluid
component
of the liquid is zero in the vortex frame).

Since the
Eq.(\ref{IordanskiiForce}) was obtained using the asymptotic befavior of
the flow
field ${\bf v}_s$, which induces the same effective metric as the metric around
the spinning string, one can apply this result directly to the spinning string.
The asymmetric cross-section of the scattering of relativistic particles on the
spinning string is given by Eq.(\ref{sigmaPerpVortex}). This means that in the
presence of the momentum of matter the spinning cosmic string experiences the
kind of the lifting  force, which coresponds to the Iordanskii force in
superfluids.  This force can be obtained by relativistic generalization of the
Eq.(\ref{IordanskiiForce}).  The momentum density
${\bf P}_n$ of quasiparticles should be substituted by the component
$T_0^i$ of the energy-momentum tensor. As a result, for  2+1
space-time and for small energy $E$, which corresponds to the low temperature
$T$ of the matter, the Iordanskii force on spinning string moving
with respect to the matter is
\begin{equation}
F_{\rm Iordanskii}^{\alpha}=8\pi JG  e^{\alpha\beta\gamma} u_\beta u_\mu
T^\mu_\gamma~.
\label{RelativisticIordanskiiForce}
\end{equation}
Here $u_\alpha$ is the 3-velocity of the string and $T^\mu_\gamma$ is the
asymptotic value of the energy-momentum tensor of the matter at the spot of the
string. Using the Einstein equations one can rewrite this as
\begin{equation}
F_{\rm Iordanskii}^{\alpha}=  J   e^{\alpha\beta\gamma} u_\beta u_\mu
R^\mu_\gamma~,
\label{CurvatureIordanskiiForce}
\end{equation}
where $R^\mu_\gamma$ is the Riemannian curvature at the position of the string.
This corresponds to the force acting on particle   with the spin $J$ from the
gravitational field due to interaction of the spin with the Riemann tensor
\cite{SpinRiemann}.

{\it Conclusion.}

There is an analogy between the asymptotic velocity field far from the vortex
core in superfluids and the gravitational field induced by the spinning cosmic
string. As a result both systems experience the gravitational Aharonov-Bohm
effect, which in particular leads to the Irodanskii force acting on the
vortex (the spinning string), when it moves with respect to the heat bath of
quasiparticles (the matter).

Iordanskii force has been experimentally identified in the rotating superfluid
$^3$He-B. According to the theory for the  transport of vortices in
$^3$He-B, the Iordanskii force completely determines the mutual friction
parameter
$d_\perp
\approx - \rho _n/\rho$ at low $T$ \cite{KopninVolovik}, where $\rho$ is the
total density of the liquid. This is in accordance with the experimental data,
which show that
$d_\perp$ does approach its negative asymptote at low $T$ \cite{Bevan}. At
higher $T$ another topological force, which comes from the spectral flow of the
fermion zero modes in the vortex core
\cite{BevanNature,AxialAnomaly}, becomes dominating leading to the sign
reversal
of $d_\perp$.  The observed negative sign of
$d_\perp$  at low $T$   provides the experimental verification of
the analog of the gravitational Aharonov-Bohm effect on spinning cosmic
string.

I thank Pawel Mazur and  Edouard Sonin  for illuminating discussions and
Bjorn Jensen for sending copies of his papers. This work was supported by the
Russian Foundation for Fundamental Research grant No. 96-02-16072, by the RAS
program ``Statistical Physics'' and by  European Science Foundation
network on ``Topological Defects in Cosmology and Condensed Matter
Physics''.

\end{document}